\documentclass[aps,pre,twocolumn,superscriptaddress,showpacs,showkeys]{revtex4-1}
\usepackage{color}
\usepackage{amssymb}
\usepackage{amsmath}
\usepackage{bm}
\usepackage{graphicx}
\usepackage{epstopdf}
\usepackage{newtxtext,newtxmath}
\usepackage{xfrac}
\usepackage{amsfonts}

\newcommand{\ki}{{k_{\rm new}}}

\newcommand{\Bmean}{{\langle\mathbb{B}\rangle}}
\newcommand{\Vmean}{{\langle\mathbb{V}\rangle}}
\newcommand{\Cmean}{{\langle\mathbb{C}\rangle}}

\begin{document}
\setcounter{page}{1}
\title{Understanding the temporal pattern of spreading in heterogeneous networks: Theory of the mean infection time}
\author{Mi Jin  \surname{Lee}}
\author{Deok-Sun \surname{Lee}}
\email{deoksun.lee@inha.ac.kr}
\affiliation{Department of Physics, Inha University, Incheon 22212, Korea}

\begin{abstract}
For a reliable prediction of an epidemic or information spreading pattern in complex systems,  well-defined measures are essential.   In the susceptible-infected model on heterogeneous networks, the cluster of infected nodes in the intermediate-time regime  exhibits too large fluctuation in size to use its mean size as a representative value. The cluster size follows quite a broad distribution, which is shown to be derived from the  variation of the cluster size  with the time when a hub node was first infected. On the contrary, the distribution of the  time taken to infect a given number of nodes is well concentrated at its mean, suggesting the mean infection time is a better measure. We 
show that the mean infection time can be evaluated by using the scaling behaviors of the boundary area of the infected cluster and use it to find a non-exponential but algebraic spreading phase in the intermediate stage on strongly heterogeneous networks. Such slow spreading originates in only small-degree nodes left susceptible, while most hub nodes are already infected in the early exponential-spreading stage. Our results offer a way to detour around large statistical fluctuations and  quantify reliably the temporal pattern of spread under structural heterogeneity.
\end{abstract}

\date{\today}
\maketitle 

\section{introduction}
\label{sec:intro}

Heterogeneity of the connectivity of elements in complex systems~\cite{barabasi99} leads to peculiar dynamic behaviors,  including large connected components formed with a small number of links~\cite{albert00},  the onset of a global epidemic~\cite{sis2} or synchronization~\cite{PhysRevE.70.026116} at all positive  interaction strengths, and a novel singularity of the free energy in the Ising and the Potts model~\cite{PhysRevE.66.016104,PhysRevE.66.036140,lee04}.  Different dynamic influences of  nodes essentially determined by their degrees (numbers of connected nodes) have been shown to underlie such anomalous emergent behaviors by extensive studies on the structure and dynamics of complex networks~\cite{Boccaletti:2006aa,dorogovtsev08,barrat2008dynamical}. 

This  advancement in our understanding is, however, restricted to  the equilibrium or stationary state. In reality, taking quick action before reaching the stationary state is necessary to control, e.g.,  the spread of a life-threatening virus or the word about marketed products. Despite such importance, theoretical understanding of the nonstationary state is far from complete. This is partly because of the time variation of relevant variables and having neither small nor large  order parameters in the intermediate-time regime, defying analytic approaches based on approximations which are  valid in the early- or late-time regime. Moreover, the speed of epidemic spreading shows a large statistical fluctuation, presumably due to the heterogeneity of the infection seed's degree and the stochasticity of infection trajectories~\cite{PhysRevLett.92.178701,*BARTHELEMY2005275,Volz2008,*Miller2011}. As we will address here, the  epidemic size at a given time suffers from such a large fluctuation that it is disqualifed from being a reliable measure, particularly in the intermediate stage of spreading. Therefore a new reliable measure and its theory are required to predict and control efficiently disease and information spreading in real-world complex systems.

Here we pay attention to epidemic spreading processes, for which various approaches have been proposed, such as pair approximation~\cite{PhysRevX.3.021004,*Mata_2013,*TRAPMAN2007464}, branching process approach~\cite{PhysRevLett.96.038702}, the moment closure method~\cite{Krishnarajah2005,*BAUCH2005217,*PEYRARD2008383}, and message passing~\cite{PhysRevE.82.016101,*PhysRevLett.112.118701}. We study the simplest model, the susceptible-infected (SI) model, to investigate thoroughly the statistical fluctuations appearing in the temporal pattern of spreading and provide the theory for an alternative reliable measure. To quantify  fluctuation, the distribution of the number of infected nodes $I$ at a given time $t$ is measured, which turns out to be so broad that the mean $\langle I\rangle_t$ loses its representativeness in  the intermediate-time regime on heterogeneous networks.  
We show analytically that the asymptotic behavior of the distribution  is derived from  the dependence of the epidemic size $I$ on the time when a hub, defined here as a node with a degree larger than 30\% of the maximum degree, is first infected. In contrast, the distribution of the time $t$ taken to infect a given number $I$ of nodes is well concentrated at its mean. This suggests that  the mean infection time $\langle t\rangle_I$ can be a measure for the reliable description of the spreading phenomena.  We construct a theory to evaluate the mean infection time, which leads us to discover, for strongly heterogeneous networks, the  algebraic relation between $\langle t\rangle_I$ and $I$
 in the intermediate stage contrasted with the well-known exponential spreading in the early stage.  The origin of the algebraic-spreading phase is investigated,
which helps us understand the temporal complexity  derived from the structural heterogeneity in various complex systems.

In Sec.~\ref{sec:model}, the SI model and the model networks are described along with their numerical  implementation. We compare and analyze the fluctuations of the number of infected nodes and the infection time in Sec.~\ref{sec:fluctuation}.  Our theory for the mean infection time is presented in Sec.~\ref{sec:meaninfectiontime}. We summarize and discuss the results in Sec.~\ref{sec:conclusion}.

\section{Model}
\label{sec:model}

%%%%% Figure 1: Fluctuation %%%%%%%%%%%
\begin{figure}
\includegraphics[width=\columnwidth]{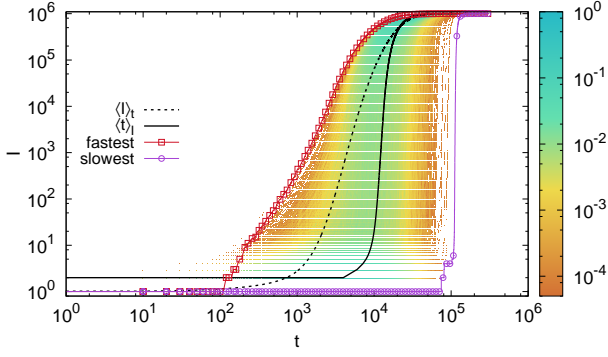}
\caption{
%Different temporal patterns of viral disease spreading 
%(a) The cumulative number of  patients versus the number of days from each outbreak of Ebola in 2014. 
%(b) The same plots as in (a) for MERS. 
%(c) 
Fluctuation in the spreading of infection on SF networks.
The fraction of simulation runs of the SI model yielding $I$ infected nodes at time $t$ is color-coded. The SI model with infection rate $\lambda = 10^{-4}$ is simulated $100$ times in each of $200$ SF networks of $N=10^6$ nodes, $L=2\times 10^6$ links (the mean degree $\langle k\rangle=4$), and degree exponent $\gamma=2.75$. 
Shown are $\langle I\rangle_t$, $\langle t\rangle_I$,  and the  fastest and slowest spreading, taking the shortest and longest time to infect $I_0 = 100$ nodes, respectively.}
\label{fig:fluctuation}
\end{figure}
%%%%%%%%%%%%%%%%%%%%%%%%%%%%%

%%%%% Model
We consider the SI model on  scale-free (SF) networks of $N$ nodes and $L$ undirected links, displaying a  power-law degree distribution $P_{\rm degree}(k)\sim k^{-\gamma}$ for large $k$, with $\gamma$ being the degree exponent.
In simulations, we use the uncorrelated configuration model~\cite{ucm} to construct the SF networks, in which each node is assigned $k$ link stubs with degree $k$ selected as described below such that its distribution takes a prescribed power-law form, and then those stubs are randomly paired until no single or pair of stubs is left. Finally, an unpaired stub, multiple links, and self-loops are removed, the numbers of which are  negligible in all the considered cases. For given $N, L,$ and $\gamma$, the degree $k_i$ of node $i$ is given by the integer part of a real-valued random number $r$ from a distribution $p(r) = p_1 r^{-\gamma}$ for $r_0 <r<  \sqrt{N}+1$ with $p_1$ being a normalization constant and  $r_0$ determined such that the resultant mean degree $\langle k\rangle = N^{-1}\sum_i k_i$ is equal to $2L/N$~\cite{hk}. The degree cannot be larger than $\sqrt{N}$, a constraint imposed to remove the degree-degree correlation of neighboring nodes, and actually the maximum degree $k_{\rm max}$  behaves as $k_{\rm max}\simeq \sqrt{N}$ for $2<\gamma<3$ and $k_{\rm max}\sim N^{1\over \gamma-1}$~\cite{ucm}.

In the SI model,  the state $x_i$ of  node $i$ is either susceptible ($x_i=0$) or infected ($x_i=1$). A susceptible node becomes infected with rate $\lambda$ by each of its infected neighbors while the transition from infected to susceptible is disallowed. 
We run the simulation of the SI model  by asynchronous updating~\cite{portergleeson2016} as follows.
%\begin{enumerate}
(i) At the initial stage ($t=0$), a randomly selected node is infected.
(ii) At each time $t$, we count the number $\mathbb{B}$ of links having a susceptible node at one end and an infected node at the other end. And we select randomly one such link and infect the susceptible node with probability $\lambda$. This is repeated $\mathbb{B}$ times to move to the next time step $t+1$. 
(iii) Repeat step (ii) until all the nodes are infected.
%\end{enumerate}

%%%%%%%%% Figure 2: Distribution
\begin{figure}
\includegraphics[width=\columnwidth]{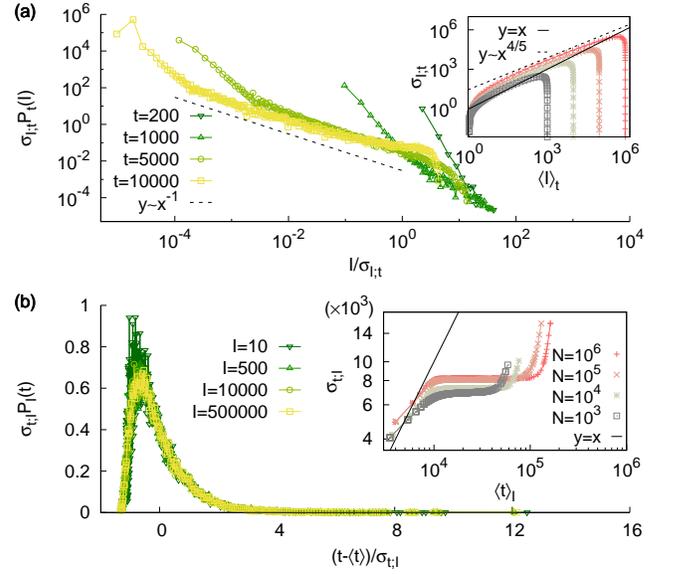}
\caption{Statistics of the infection spreading in the SI model on SF networks with $\gamma=2.75$. 
(a) The probability distribution $P_t (I)$ of $I$ at time $t$  for the network of $N=10^6$.  $\sigma_{I;t}$ is its standard deviation. The dashed line $y=0.003x^{-1}$ is shown as a guide. Inset: $\sigma_{I;t}$ versus the mean $\langle I\rangle_t$ for different $N$. The lines $y=x$ (solid) and $y = 30 x^{4/5}$ (dashed) are shown.   
(b) The distribution $P_I (t)$ of the time $t$ taken to infect $I$ nodes. $\langle t\rangle_I$ and $\sigma_{t;I}$ are its mean and standard deviation. Inset: $\sigma_{t;I}$ versus $\langle t\rangle_I$. The line $y=x$ (solid) is shown as a guide.}
\label{fig:distribution}
\end{figure}
%%%%%%%%%%%%%%%%%%%%%%%%%%%%%

%%%%% Fluctuations
\section{Large fluctuation of the number of infected nodes and its origin}
%\Corr{\section{Fluctuations}}
\label{sec:fluctuation}

Simulation data for the number of infected nodes $I = \sum_{i=1}^N \delta_{x_i,1}$ are scattered in the $(t, I)$ plane to an extent varying with $\gamma$ except for  quite small or large $t$. See Fig.~\ref{fig:fluctuation} for $\gamma=2.75$. See also Fig.~\ref{fig:othernet} in Appendix~\ref{app:hub} for other $\gamma$.

To quantify such a fluctuation, we measure the number of infected nodes at time $t$, which is  found in the intermediate-time regime to follow a power law
\begin{equation}
P_t(I) \sim I^{-\eta}
\label{eq:pit}
\end{equation}
over  a wide range of $I$ with the exponent $\eta\simeq 1$ [Fig.~\ref{fig:distribution}(a)].  The standard deviation is mostly not smaller than the mean $\langle I\rangle_t=\sum_I IP_t(I)$ scaling almost linearly in the time period showing $1\ll \langle I\rangle_t\ll N$ which we refer to as the intermediate-time regime.  With such a large fluctuation,  $\langle I\rangle_t$ cannot be a representative value of $I$. For instance, the probability to observe $I$ larger than $\langle I\rangle_t$ is only $0.15$ at $t=10^4$ [see Figs.~\ref{fig:fluctuation} and~\ref{fig:distribution}(a)]. 

In striking contrast, the distribution $P_I(t)$ of the time $t$ taken to infect $I$ nodes is well concentrated at its mean $\langle t\rangle_I = \int dt \, t\,  P_I (t)$ [Fig.~\ref{fig:distribution}(b)]. The standard deviation remains far smaller than the mean unless $I$ is too small, demonstrating that $\langle t\rangle_I$ is a well-defined measure.  Note that the line representing $\langle t\rangle_I$ is in the middle of the region showing high probability in the $(t,I)$ plane (Fig.~\ref{fig:fluctuation}), supporting its representativeness. Therefore one should refer to how long it will take to infect a given number of nodes, rather than how many will be infected at a given time, in describing and predicting the pattern of spreading over heterogeneous contact networks. The difference between the two mean values grows with $N$ in SF networks (see Appendix~\ref{app:ratio}).
%%%%% Figure 3: t_hub %%%%%%%%%%%
\begin{figure}
\includegraphics[width=\columnwidth]{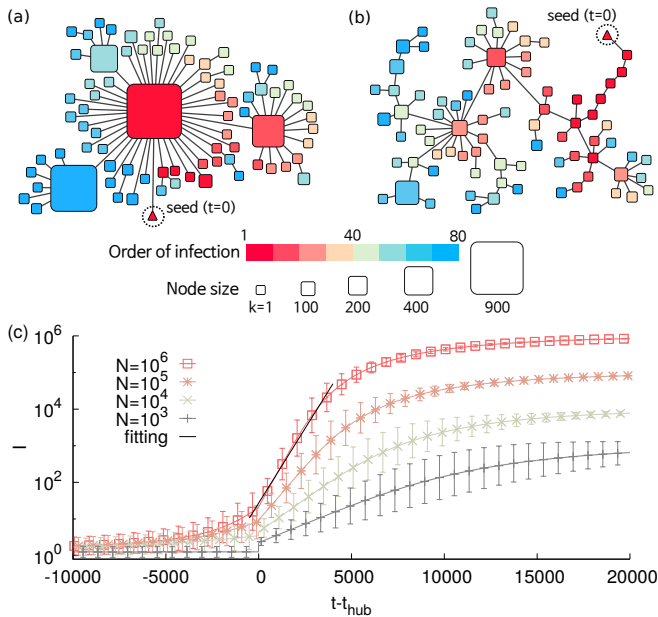}
\caption{
Role of infecting hubs in infection spreading on a SF network with $\gamma=2.75$.  
The cluster of the first $80$ infected nodes appears (a) at the observation time $t=550$ (fastest spreading) or (b) at $t=110870$ (slowest spreading). Node size and color  represent the degree and the infection order of each node. 
(c) Plot of $I$ versus the difference between $t$ and the hub-infection time $t_{\rm hub}$. The mean and the standard deviation  are shown, along with a solid line fitting Eq.~(\ref{eq:logIthub}) to the data.  
 }
\label{fig:thub}
\end{figure}
%%%%%%%%%%%%%%%%%%%%%%%%%%%%%

Before addressing the theory for the mean infection time, let us consider why $P_t(I)$ decays so slow in Eq.~(\ref{eq:pit}).
Hubs are abundant in SF networks and their infection should play a role in speeding up or slowing down spreading~\cite{PhysRevLett.92.178701,*BARTHELEMY2005275}.    As seen in Figs.~\ref{fig:thub}(a) and ~\ref{fig:thub}(b),  the cluster of the first $80$ infected nodes in the  fastest spreading has many hub nodes infected very early, while that from the slowest  spreading has only small-degree nodes infected early and hub nodes infected late. This suggests that whether and when hubs are infected determine the growth of the infected cluster.  
To check this, we measure the hub-infection time  $t_{\rm hub}$, the earliest time when any hub is infected. Using different criteria for hubs does not change the results qualitatively. When $I$ is plotted as a function of $t-t_{\rm hub}$ [Fig.~\ref{fig:thub}(c)], its statistical fluctuation is significantly reduced in comparison to the large fluctuation for given $t$ shown in Fig.~\ref{fig:fluctuation}. Moreover, $I$ grows abruptly for $0\lesssim t-t_{\rm hub} \lesssim \Delta$, with $\Delta$ being a constant; For example, $\Delta \simeq 5000$ for $\gamma=2.75$ and $N=10^6$. This demonstrates that the global spreading can occur when hubs are infected. The number of infected nodes at time $t$  satisfies the relation
\begin{equation}
\langle \ln  I\rangle \simeq   a_0 + a_1 \, (t-t_{\rm hub})
 \label{eq:logIthub}
\end{equation}
for $0\lesssim t-t_{\rm hub} \lesssim\Delta$ with $a_0$ and $a_1$ positive constants. Given the small fluctuation of $\ln I$ with respect to its mean in Eq.~(\ref{eq:logIthub}) for given $t$ and $t_{\rm hub}$ and the observation that the probability distribution $P(t_{\rm hub})$  is almost constant $P_0$ for $|t_{\rm hub}-t|\lesssim \Delta$ (see Appendix~\ref{app:hub}), we obtain $P_t(I)$ from Eq.~(\ref{eq:logIthub}) as 
\begin{align}
P_t(I) \sim P_0 \left| {d t_{\rm hub} \over d\, I}\right| \sim {P_0/a_1 \over I},
\label{eq:ptithub}
\end{align}
which agrees with Eq.~(\ref{eq:pit}).
This finding provides a guideline for the epidemic-size distribution; $P_t(I)$ different from Eq.~(\ref{eq:pit}) implies a relation other than Eq.~(\ref{eq:logIthub}).
$t_{\rm hub}$ is expected to depend on the network characteristics of the initially infected node (seed)~\cite{PhysRevLett.92.178701,*BARTHELEMY2005275} and also on the specific realization of spreading in the early stage. We find  both the degree of the seed and its shortest distance to a hub significantly correlated with $t_{\rm hub}$ (see Fig.~\ref{fig:thubdeterminants} in Appendix~\ref{app:hub}). In practically controlling the epidemic spreading, various factors can be influential, such as the $k$ core~\cite{Kitsak:2010aa},  and should be considered when designing efficient intervention strategies~\cite{Massaro:2018aa} and identifying superspreaders and superblockers~\cite{Morone:2015aa, PhysRevE.95.012318}.

\section{Analytic approach to the mean infection time}
\label{sec:meaninfectiontime}

According to the conventional mean-field theory applied to heterogeneous networks~\cite{sis1, barrat2008dynamical},  the  probability of a susceptible node to be infected per unit time interval is proportional to its degree and  the probability of encountering an infected neighbor. The latter probability is assumed to be a function of time and is solved in a self-consistent way to reveal exponential growth and saturation of the number of infected nodes in the early- and late-time regimes, respectively~\cite{sis2,barrat2008dynamical,PhysRevLett.92.178701,RevModPhys.87.925}. However, in the intermediate-time regime, large fluctuations prevent us from referring to time-dependent functions.

To construct a theory for the mean infection time, let us first consider the time $\tau$ taken to newly infect a susceptible node,  the average of which will be identified with  ${d\langle t\rangle_I \over dI}$.  As infection spreads along the links connecting an infected node and a susceptible node, the total number  $\mathbb{B}$ of such links, which is counted during the simulation of the SI model as in Sec.~\ref{sec:model} and we call {\it boundary} links, essentially determines $\tau$.  The boundary links were also used to formulate the uniform mean-field framework for time-dependent quantities in~\cite{PhysRevE.96.052314}.  If a cluster of infected nodes has $\mathbb{B}$ boundary links,  a newly infected node will first appear at a time between $\tau$ and $\tau+d\tau$ with probability  $P_{\mathbb{B}} (\tau) d\tau = \lambda \, d\tau  \, \mathbb{B} e^{-\lambda \tau \mathbb{B}}$. Given $I$ infected nodes, the fluctuation of $\mathbb{B}$ is insignificant unless $I$ is too small (see Fig.~\ref{fig:sigmaB} in Appendix~\ref{app:pdki}), allowing us to use  the mean $\Bmean_I$. Therefore,  we evaluate the mean time $\langle \tau\rangle_I$ taken to infect one more node given $I$ infected ones as 
\begin{align}
{d \langle t\rangle_I \over dI } &= \langle \tau \rangle_I \simeq  \int_0^\infty d\tau P_{\Bmean_I }(\tau) \, \tau \nonumber\\
&=  {1\over \lambda \, \Bmean_I} = {1\over \lambda (\Vmean_I - \Cmean_I)},
\label{eq:dtdi}
\end{align}
where  we introduced  
\begin{align} 
\Vmean_I &\equiv \sum_{i,j} A_{ij} \langle \delta_{x_i,1}\rangle_I = \sum_{i} k_i \langle \delta_{x_i,1}\rangle_I = \sum_{I'=0}^{I-1} \langle \ki\rangle_{I'}, \nonumber\\
\Cmean_I &\equiv \sum_{i,j} A_{ij} \langle \delta_{x_i,1} \delta_{x_j,1}\rangle_I,
\label{eq:VC}
\end{align}
with $A$ being the adjacency matrix, which is symmetric, and $\langle \ki\rangle_I$ being the expected degree of the newly infected node given $I$ infected nodes or, equivalently, the $(I+1)$th infected node.  
$\mathbb{V}$ and $\mathbb{C}$ are the link-based volume of the whole and the internal part of  the infected cluster; 
%They are the sum of the number of all links incident to each infected node, and of the links incident from other infected nodes to  each infected node, respectively. 
$\mathbb{V}$ ($\mathbb{C}$) is the sum of the number of all (infected) neighbors of all infected nodes. In this sense,  $\mathbb{B} = \sum_{i,j} A_{ij} \delta_{x_i,1}\delta_{x_j,0} = \mathbb{V}-\mathbb{C}$ can be considered the boundary area of the cluster. A similar link-based approach was taken in establishing nonlinear differential equations for time-dependent variables~\cite{Volz2008,*Miller2011,C.:2012aa}.

%%%%% Figure 4: algebraic growth  %%%%%%%%%%%
\begin{figure}
\includegraphics[width=\columnwidth]{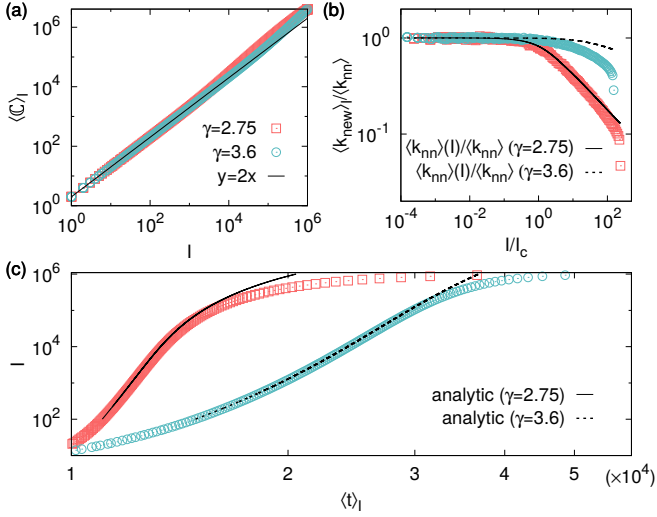}
\caption{The volume of the infected cluster and the mean infection time $\langle t\rangle_I$ for $N=10^6$. 
(a) The internal volume $\Cmean_I$ versus $I$. The approximation in Eq.~(\ref{eq:Khat}) is shown as a guide.
(b) The degree $\langle \ki\rangle_I$ of the newly infected node given $I$ infected nodes. Its cumulative sum gives the whole volume $\Vmean_I$ as in Eq.~(\ref{eq:VmeanI}). The lines represent $\langle k_{\rm nn}\rangle(I)/\langle k_{\rm nn}\rangle$ computed by using Eq.~(\ref{eq:ki1}) with $P_{\rm degree}(k)$ from the simulations used. 
(c) Plots of $I$ versus $\langle t\rangle_I$ from simulations  (points) and from the solutions (lines) to Eq.~(\ref{eq:dtdi}) with Eqs.~(\ref{eq:Khat}) and (\ref{eq:VmeanI}) and the initial condition $I_0 = 100$ and $t_0 = \langle t\rangle_{I_0}$. 
}
\label{fig:algebraic}
\end{figure}
%%%%%%%%%%%%%%%%%%%%%%%%%%%%%

To complete and solve Eq.~(\ref{eq:dtdi}), the $I$ dependence of $\Vmean_I$ and $\Cmean_I$ should be known. 
%The infection  of a new node tends to increase $\mathbb{C}$ by $2$, as the link used to transmit this infection is added to the internal part of the cluster, 
When a node $\ell$ is newly infected, $\mathbb{C}$ is increased by twice the number of its previously infected neighbors, $2\sum_j A_{\ell j} \delta_{x_j,1}$. When $I$ is not so large,  the node $\ell$ is very likely to have just one infected neighbor, without forming a loop in the infected cluster as supported by its tree structure, as seen in Figs.~\ref{fig:thub}(a) and~\ref{fig:thub}(b)~\cite{PhysRevE.93.052310}. $\mathbb{C}$ is increased by $2$ whenever a newly infected node appears, resulting in
\begin{equation}
\Cmean_I \simeq 2I.
\label{eq:Khat}
\end{equation}
It is valid for a wide range of $I$  [Fig.~\ref{fig:algebraic}(a)], except for the large-$I$ region  where a newly infected node can have more than one infected neighbor, forming loops in the infected cluster.

Next, we consider the degree of the $(I+1)$th infected node. Before its infection, the node was susceptible and connected to one of the  $I$ previously infected nodes.  Let us assume that every link from the susceptible nodes is equally likely to be heading to one of the infected nodes. Then the probability $r_I(k)$ that a susceptible neighbor of the $I$ infected nodes has degree $k$ can be approximated as $r_I (k) \simeq k n(k|I)/\sum_{k'} k' n(k'|I)\simeq k n(k|I)/(2L)$, where  $n(k|I)$ is the expected number of susceptible nodes with degree $k$ given $I$ infected nodes. We also assumed $\langle \mathbb{V}\rangle_I\ll 2L$ in the relation $\sum_{k'} k' n(k'|I)=2L-\langle \mathbb{V}\rangle_I\simeq 2L$, which is valid in the intermediate stage. The decrease in the number of susceptible nodes of degree $k$, $n(k|I) - n(k|I+1)$, is equal to $r_I(k)$, giving 
\begin{equation}
n(k|I+1) \simeq \left(1- {k\over 2L}\right) n(k|I).
\label{eq:recursion}
\end{equation}
Here $1- k/(2L)$ is the probability that  any link of a susceptible node of degree $k$ is not used to transmit infection when a newly infected node appears. From Eq.~(\ref{eq:recursion}), one obtains $n(k|I)\simeq  n(k|I=0) e^{-{k I\over 2L}}$. The expected degree of the $(I+1)$th infected node $\langle \ki\rangle_I = \sum_k k r_I(k)$ is evaluated as
\begin{align}
\langle \ki\rangle_I \simeq  \langle k_{\rm nn}\rangle(I)\equiv  \frac{\sum_{k=1}^{k_{\rm max}} k^2 P_{\rm degree}(k) e^{-{k \, I\over 2L}}}{\sum_{k=1}^{k_{\rm max}} kP_{\rm degree}(k) e^{-{k\, I\over 2L}}},
\label{eq:ki1}
\end{align}
where we defined $\langle k_{\rm nn}\rangle(I)$, which is reduced to the mean degree of a node's neighboring node $\langle k_{\rm nn}\rangle = \sum_k k^2 P_{\rm degree} (k)/\sum_k k P_{\rm degree}(k)$ for $I=0$. Notice that $\langle k_{\rm nn}\rangle(I)$ is computed by using the degree distribution of the underlying network. 

Simulations  support the agreement between $\langle \ki\rangle_I$ and $\langle k_{\rm nn}\rangle(I)$ [Fig.~\ref{fig:algebraic}(b)]. $\langle \ki \rangle_I$ is  constant for small $I$ but decreases with $I$ for large $I$, particularly in SF networks with small $\gamma$. A similar decrease in the degree of newly infected nodes with time was noted in ~\cite{PhysRevLett.92.178701,*BARTHELEMY2005275}. However, its functional behavior remains unknown, which should be understood for the theory of the mean infection time. The exponential term $e^{-{k I\over 2L}}$ in Eq.~(\ref{eq:ki1}) is the key.
When  $I$ is so small that $ k_{\rm max}\, I/(2L)\ll 1$ or $I\ll I_{\rm c}$, with 
\begin{equation}
I_{\rm c} \equiv {2L \over k_{\rm max}},
\label{eq:I1}
\end{equation}
the exponential term is close to $1$  for all $k\leq k_{\rm max}$,  and  thus $\langle \ki\rangle(I)\simeq \langle k_{\rm{nn}}\rangle(I=0)=\langle k_{\rm nn}\rangle$.  If $I\gg I_{\rm c}$, $e^{-{k I\over 2L}}$ will be quite small for $k\gg \tilde{k}(I)\equiv 2L/I$, meaning that susceptible nodes with a degree larger than  $\tilde{k}(I)$ are rarely seen, as they are already infected,  causing $\langle k_{\rm nn}\rangle (I)$ to decrease with $I$. In the configuration-model SF networks~\cite{ucm}, $k_{\rm max}\sim N^{1 \over 2}$ for $2<\gamma <3$, and $k_{\rm max}\sim N^{1 \over (\gamma-1)}$ for $\gamma >3$. Therefore the intermediate stage of infection is divided into two regimes: $1\ll I\ll I_c$ and $I_c\ll I\ll N$. The decay of $\langle k_{\rm nn}\rangle (I)$ with $I$ for $I\gg I_c$ is significant in SF networks with $2<\gamma<3$, for which $\langle k_{\rm nn}\rangle (I)$ diverges with $\min\{k_{\rm max},\tilde{k}(I)\}$. As $N\to\infty$, $\langle k_{\rm nn}\rangle(I)\simeq \langle k_{\rm nn}\rangle\sim k_{\rm max}^{3-\gamma}$ for $I\ll I_{\rm c}$, and $\langle k_{\rm nn}\rangle(I) \sim \tilde{k}(I)^{3-\gamma}\sim I^{-(3-\gamma)}$ for  $I\gg I_{\rm c}$ (see Appendix~\ref{app:asymptotic}).

Solving Eq.~(\ref{eq:dtdi}) by using the approximation for $\Vmean_I$,
\begin{equation}
\Vmean_I \simeq \sum_{I'=0}^{I-1} \langle k_{\rm nn}\rangle(I'),
\label{eq:VmeanI}
\end{equation}
which behaves as $\langle k_{\rm nn}\rangle I$ for $I\ll I_c$ and $I^{\, \gamma-2}$ for $I\gg I_c$, and  using Eq.~(\ref{eq:Khat}) for $\Cmean_I$,  one obtains $\langle t\rangle_I$ from $P_{\rm degree} (k)$ of the substrate networks. In Fig.~\ref{fig:algebraic}(c), the simulation data agree with this solution in the intermediate stage. 

The analytic solution to $\langle t\rangle_I$  reveals a crossover around $I_c$ from the exponential- to algebraic-spreading phase in SF networks of $2<\gamma<3$.
%to see its large $N$ behaviors. 
For $1 \ll I\ll I_{\rm c}$,  $\langle k_{\rm nn}\rangle(I)$ is fixed at $\langle k_{\rm nn}\rangle$,  yielding the exponential spreading
\begin{equation}
I \simeq I_0\, e^{(\langle k_{\rm nn}\rangle -2) \lambda \, (\langle t\rangle_I-t_0)},
\label{eq:Iearly}
\end{equation}
where $I_0$ is a constant larger than $1$ but much smaller than $I_{\rm c}$ and $t_0 = \langle t\rangle_{I_0}$. 
In SF networks with $\gamma>3$, $\langle \ki \rangle_I$ decreases very weakly with $I$ for $I_c\ll I\ll N$, and therefore, Eq.~(\ref{eq:Iearly}) is valid approximately for $1\ll I\ll N$. On the contrary, in SF networks with $2 < \gamma<3$, the sub-linear growth of $\Vmean_I$ for $I\gg I_{\rm c}$ leads to  
\begin{align}
I &\simeq  N  \, a \,  (\lambda \langle t\rangle_I)^{1/(3-\gamma)},
\label{eq:I_I1_g}
\end{align}
with the coefficient $a \equiv \langle k\rangle [(\langle k_{\rm nn}\rangle/k_{\rm max}^{3-\gamma}) (3-\gamma)\Gamma(4-\gamma)/(\gamma-2)]^{1/(3-\gamma)}$ (see Appendix~\ref{app:asymptotic}). This means that  infection spreads  with time algebraically, slower than an exponential spreading. The polynomial $t$ dependence of $\langle I\rangle_t$ has been studied using the branching process approach~\cite{PhysRevLett.96.038702}.
It reflects the inequivalent chances of infection for nodes of different degrees; most hub nodes are infected for $I\ll I_c$, and only the small-degree nodes are left susceptible for $I\gg I_c$.
Knowing such crossover in the spreading speed can be helpful for designing and executing in a timely fashion an efficient strategy to intervene in the spreading process.

\section{Conclusion}
\label{sec:conclusion}

To conclude, we have shown that the infection time is well defined as a function of the number of infected nodes, enabling the reliable description and prediction of the temporal pattern of spreading  in heterogeneous networks.  The link-based volume and boundary area of the infected cluster were  investigated as a function of its size, which allowed us to see how the node degree affects the order of infection and understand the temporal complexity characterized by the algebraic spreading in the nonstationary state. In more complex spreading dynamics such as the susceptible-infected-susceptible or susceptible-infected-recovered models, the infected cluster may shrink in the bulk due to recovery as well as grow at the boundary, which could deepen our understanding of the spreading phenomena. The perspective and method presented in this work can be used in practical applications as well as in the study of various model dynamics  on heterogeneous networks.

%%%% Acknowledgment
\begin{acknowledgments}
This work was supported by National Research Foundation of Korea (NRF) grants funded by the Korean government (Grant No. 2016R1A2B4013204).
\end{acknowledgments}
%\bibliography{SI}

\appendix
\section*{Appendix}

\section{Fluctuations of the number of infected nodes and the infection time}
\label{app:std}

%%%%%%%% Figure: std/mean %%%%%%%%%%%%%%%%%
\begin{figure}
\includegraphics[width=\columnwidth]{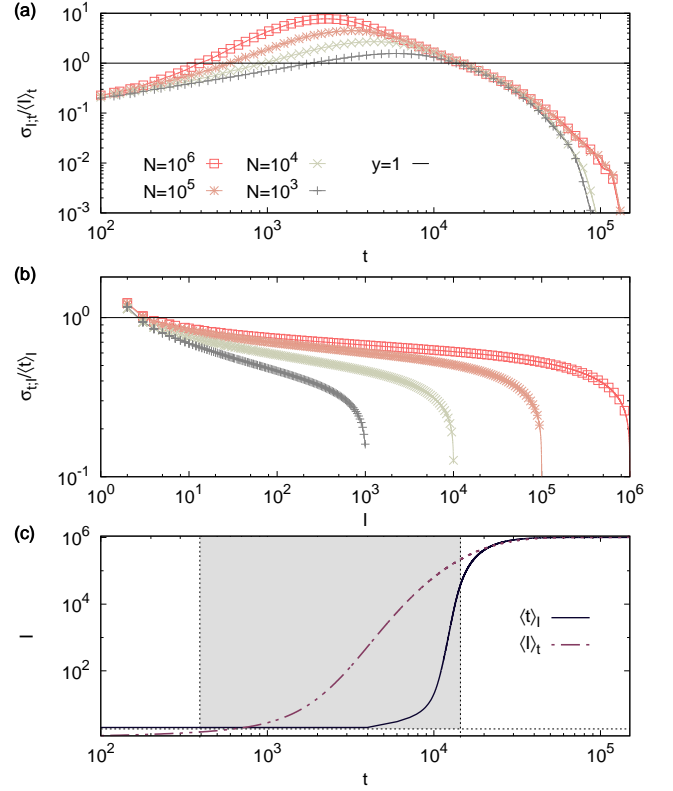}
\caption{ Relative fluctuations of the number of infected nodes and the infection time in the SI model on SF networks with $\gamma=2.75$ and different numbers of nodes $N$. 
(a) The ratio of the standard deviation $\sigma_{I;t}$ to  the mean $\langle I\rangle_t$ of the number of infected nodes at  time $t$. 
(b) The ratio of the standard deviation $\sigma_{t;I}$ to the mean $\langle t\rangle_I$ of the time to infect $I$ nodes  in the same networks as in (a).
(c) The region where  $\sigma_{t;I}<\langle t\rangle_I$ and  $\sigma_{I;t}>\langle I\rangle_t$ is shaded in the $(t, I)$ plane.
}
\label{fig:relfluct}
\end{figure}
%%%%%%%%%%%%%%%%%%%%%%%%%%%%%%%%%%

For a random variable, its standard deviation should be smaller than the mean if the mean is to be used as a representative value. We present the relative fluctuation, the  ratio of the standard deviation to the mean, of the number of infected nodes at each given time and of the time taken to infect a given number of nodes in Figs.~\ref{fig:relfluct}(a) and~\ref{fig:relfluct}(b). In SF networks with $\gamma=2.75$ and $N=10^6$, the mean number of infected nodes at a given time $\langle I\rangle_t$ is a good measure only in the early-time regime, $t\lesssim 500$, or in the late-time regime $t\gtrsim 20\,000$. In the intermediate-time regime, $500 \lesssim t \lesssim 20\,000$, the standard deviation $\sigma_{I;t}$ is not smaller than the mean $\langle I\rangle_t$. On the other hand, the mean infection time $\langle t\rangle_I$ is well defined as long as $I\gtrsim 3$. One can see a broad region in the $(t,I)$ plane where  only the mean infection time $\langle t\rangle_I$ is well defined  in Fig.~\ref{fig:relfluct}(c).

\section{Difference between $\langle I\rangle_t$ and $\langle t\rangle_I$}
\label{app:ratio}

%%%%%%%% Figure: curve diff %%%%%%%%%%%%%%%%%
\begin{figure}
\includegraphics[width=\columnwidth]{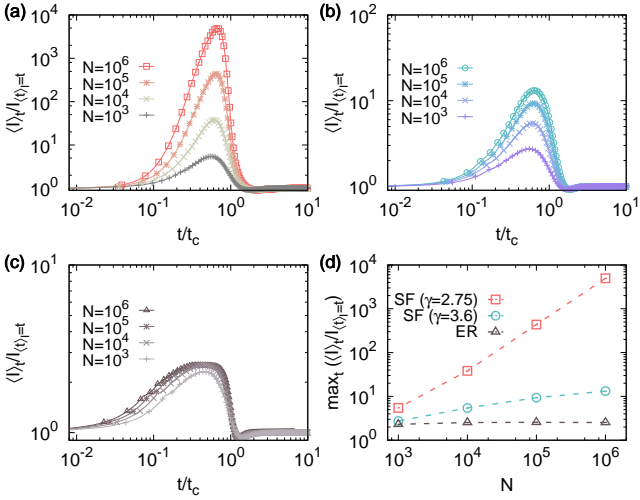}
\caption{Ratio of  the mean number of infected nodes $\langle I\rangle_t$ at time $t$ to  the value of $I$ at which $\langle t\rangle_I=t$, denoted by $I_{\langle t\rangle_I = t}$. Time $t$ in the horizontal axis is scaled by $t_c \equiv \langle t\rangle_{I_c}$ with $I_c$ in Eq.~(9).  
(a) The ratio $\langle I\rangle_t / I_{\langle t\rangle_I = t}$ in the SF networks with $\gamma=2.75$ and different numbers  of nodes $N$. 
(b) The ratio in the SF networks with $\gamma=3.6$. (c) The ratio in the ER networks. 
%It is the maximum around $t/t_1\simeq 0.5$ in all the considered cases in (a-c). 
(d) The maximum of the ratio $\max_t (\langle I\rangle_t / I_{\langle t\rangle_I = t})$ versus system size $N$. }
\label{fig:curvediff}
\end{figure}
%%%%%%%%%%%%%%%%%%%%%%%%%%%%%%%%%%

The difference between the mean number of infected nodes and the mean infection time plotted in the $(t,I)$ plane in Fig.~\ref{fig:fluctuation} appears particularly large in the  intermediate-time regime.  To see whether this  difference remains significant in the limit $N\to\infty$,  we plot the ratio of $\langle I\rangle_t$ to the value of $I$ at which $\langle t\rangle_I = t$ versus $t$ for SF networks and Erd\H{o}s-R\'{e}nyi (ER) networks in Figs.~\ref{fig:curvediff}(a)-\ref{fig:curvediff}(c). We find that the ratio is significantly larger than $1$ in the range $0.1\lesssim t/t_c \lesssim 2$, where  $t_c = \langle t\rangle_{I_c}$ is the mean time to infect $I_c$ nodes with $I_c$ in Eq.~(9). In Fig.~\ref{fig:curvediff}(d), the largest value of the ratio is shown to increase quickly with $N$ in SF networks with $\gamma=2.75$, contrary to a relatively weak or no increase in weakly heterogeneous networks. Therefore,  the difference in the two approaches relying on $\langle I\rangle_t$ and $\langle t\rangle_I$ cannot be neglected for strongly heterogeneous networks.

\section{Derivation of Equation~(3)}
\label{app:hub}

%%%%%%%% Figure: P(t03) %%%%%%%%%%%%%%%%%
\begin{figure}
\includegraphics[width=\columnwidth]{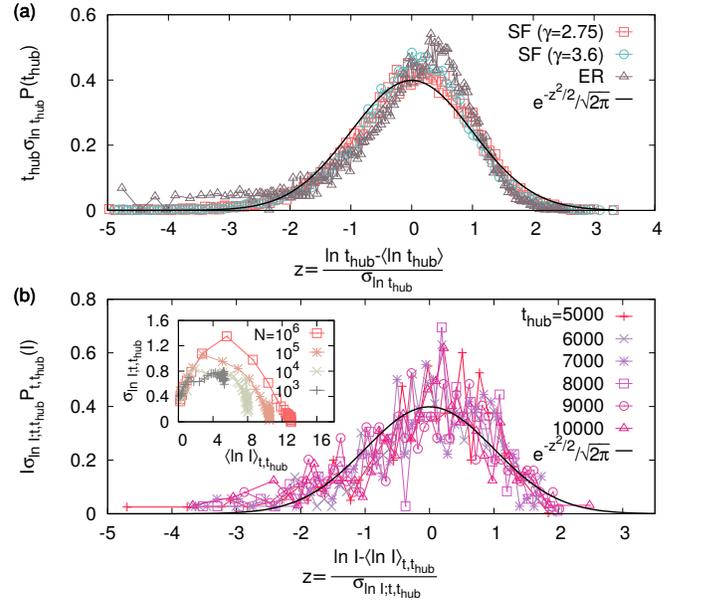}
\caption{Statistics of the hub-infection time and the number of infected nodes for a given hub-infection time. 
(a) The distributions $P(t_{\rm hub})$ of the hub infection time in the SF and ER networks of $N=10^6$.  They are fitted to log-normal distributions  with mean $\langle \ln t_{\rm hub}\rangle = 9.01, 9.74$, and $8.12$ and standard deviation $\sigma_{\ln t_{\rm hub}} = 0.78, 0.41$, and $1.22$ of $\ln t_{\rm hub}$ for SF networks with $\gamma=2.75, 3.6$ and the ER networks, respectively.  
(b) The conditional probability distribution $P_{t,t_{\rm hub}}(I)$ of the number of infected nodes at $t$ for selected hub-infection times $t_{\rm hub}$ in SF networks with $\gamma=2.75$ and $N=10^6$. They are also fitted to  log-normal distributions with the mean and standard deviation of $\ln I$ at time $t=10^4$ for given $t_{\rm hub}$ shown in the inset. 
}
\label{fig:pthub}
\end{figure}
%%%%%%%%%%%%%%%%%%%%%%%%%%%%%%%%%%

%%%%%%%%%%%%%%%%%%%%%%%%%%%%%%%%%%
\begin{figure*}	
\includegraphics[width=\textwidth]{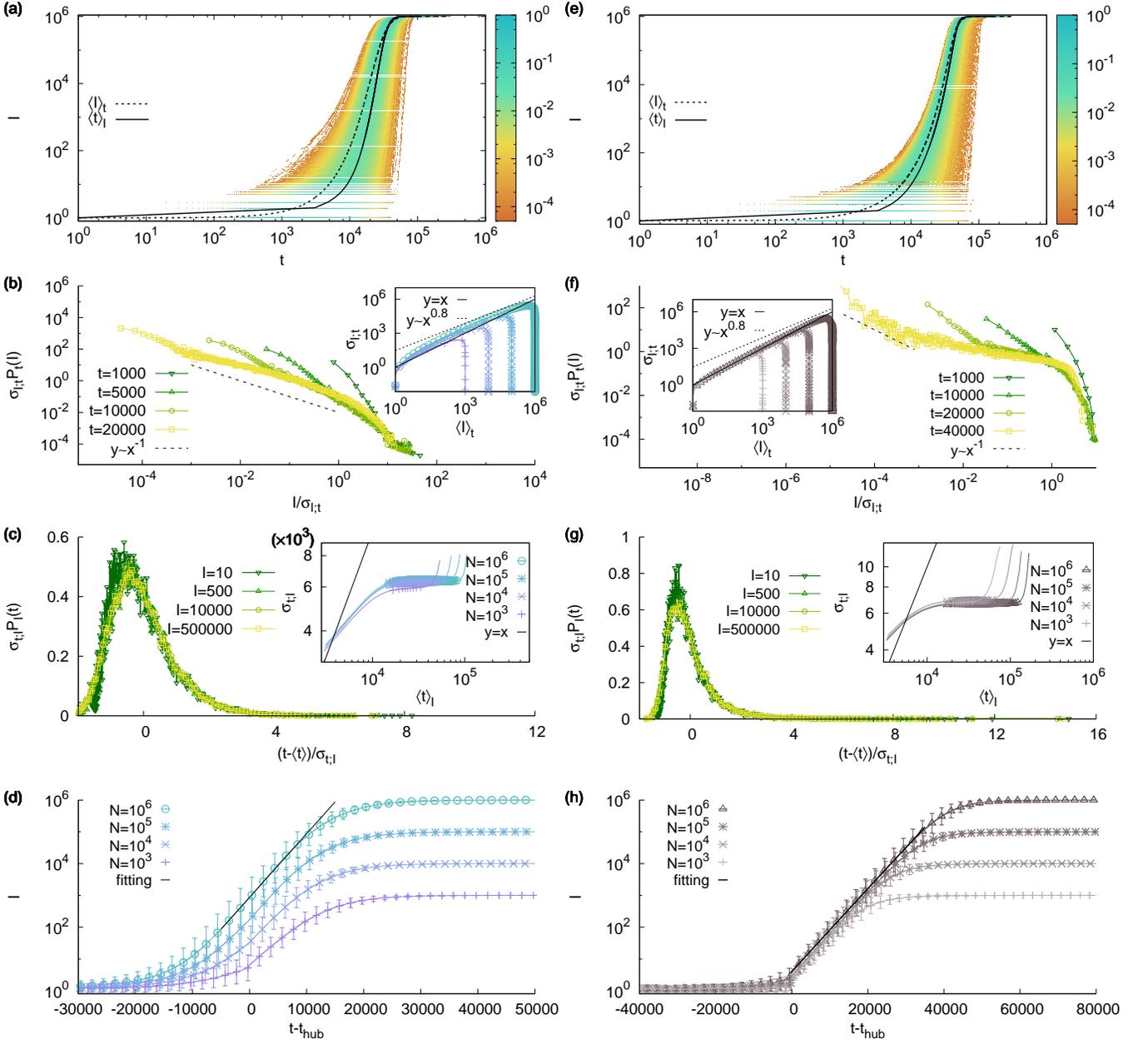}
\caption{Simulation results of the SI model on (a)-(d) SF networks with $\gamma=3.6$ and (e)-(h) ER networks.
(a) and (e) The fraction of simulations runs yielding $I$ infected nodes at time $t$ is color-coded in the $(t, I)$ plane.
(b) and (f) The probability distribution $P_t(I)$ for $N=10^6$. Inset: the standard deviation $\sigma_{I;t}$ versus the mean $\langle I\rangle_t$.
(c) and (g) The probability distribution $P_I(t)$ for $N=10^6$. Inset: the standard deviation $\sigma_{t;I}$ versus the mean $\langle t\rangle_I$.
(d) and (h) Plot of $I$ versus $t-t_{\rm hub}$. A solid line fitting Eq.~(2) is shown.
}
\label{fig:othernet}
\end{figure*}
%%%%%%%%%%%%%%%%%%%%%%%%%%%%%%%%%%

We can decompose $P_t(I)$ as
\begin{equation}
P_t(I) = \int d\, t_{\rm hub} P(t_{\rm hub}) \, P_{t,t_{\rm hub}} (I),
\label{eq:recallpti}
\end{equation}
where $P(t_{\rm hub})$ is the probability distribution of the time $t_{\rm hub}$ taken to infect a node with a degree lager than $0.3 k_{\rm max}$ and $P_{t,t_{\rm hub}}(I)$ is the conditional distribution of the number of infected nodes at time $t$ for given a $t_{\rm hub}$. Our simulation shows that $P(t_{\rm hub})$ can be fitted to a log-normal distribution as [Fig.~\ref{fig:pthub}(a)] 
\begin{equation}
P(t_{\rm hub}) \simeq {1\over t_{\rm hub}} {1\over \sqrt{2\pi \sigma_{\ln t_{\rm hub}}^2}} e^{-{(\ln t_{\rm hub} - \langle \ln t_{\rm hub}\rangle)^2\over 2\sigma_{\ln t_{\rm hub}}^2}}
\label{eq:pthub}
\end{equation}
with  $ \langle \ln t_{\rm hub} \rangle$ and $\sigma_{\ln  t_{\rm hub}}$ being the mean and the standard deviation of $\ln t_{\rm hub}$.

The conditional distribution $P_{t, t_{\rm hub}}(I)$ also takes  a log-normal form,
\begin{align}
P_{t,t_{\rm hub}}(I) &\simeq {1\over I} {1\over \sqrt{2\pi \sigma_{\ln I; t,t_{\rm hub}}^2}} e^{-{(\ln I - \langle \ln I \rangle_{t,t_{\rm hub}} )^2\over 2\sigma_{\ln I; t,t_{\rm hub}}^2}} \nonumber\\
&\simeq {1\over I} {1\over \sqrt{2\pi \sigma_{\ln I; t,t_{\rm hub}}^2}} e^{-{[t_{\rm hub}-\tilde{t}_{\rm hub}(t, I)]^2\over 2\tilde{\Delta}^2}},
\label{eq:pthubi}
\end{align}
where
\begin{equation}
\tilde{t}_{\rm hub}(t, I)=t-{{\ln I -a_0} \over a_1}, \tilde{\Delta}={\sigma_{\ln I; t,t_{\rm hub}} \over a_1}
\label{eq:tilde_td}
\end{equation}
are used, from approximating $\langle\ln I\rangle_{t, t_{\rm hub}}$ with  Eq.~(2), and $\sigma_{\ln I_{t, t_{\rm hub}}}$ is shown in the inset of Fig.~\ref{fig:pthub}(b). Our simulation results, particularly those in Figs.~3(c) and~\ref{fig:pthub}, indicate that $\tilde{\Delta}$ in Eq.~(\ref{eq:tilde_td}) is smaller than the width of $P(t_{\rm hub})$; For $\gamma = 2.75$ and $N=10^6$, we have $\tilde{\Delta}\simeq 1/a_1\simeq 500\,(\simeq \Delta /\ln N)$  while the width $w$ of the probability distribution $P(t_{\rm hub})$ is approximately $w\simeq \exp (\langle \ln t_{\rm hub}\rangle + \sigma_{\ln t_{\rm hub}})-\exp (\langle \ln t_{\rm hub}\rangle)\simeq 10^4$. Here $\Delta$ is the width of the region of $t-t_{\rm hub}$ displaying the abrupt increase in $I$ in Fig.~3(c). Therefore, inserting Eq.~(\ref{eq:pthubi}) into (\ref{eq:recallpti}), we obtain
\begin{align}
P_t (I) &\simeq  \int_{\tilde{t}_{\rm hub}(t, I)-\tilde{\Delta}}^{\tilde{t}_{\rm hub}(t, I)+\tilde{\Delta}} d\, t_{\rm hub} P(t_{\rm hub})\, P_{t,t_{\rm hub}}(I) \nonumber\\
& \simeq  P(\tilde{t}_{\rm hub}(t, I)) \int d \, t_{\rm hub}{1\over I} \, {e^{-{[t_{\rm hub}-\tilde{t}_{\rm hub}(t, I)]^2\over 2\tilde{\Delta}^2}} \over \sqrt{2\pi \sigma^2_{\ln I; t, t_{\rm hub}}}} \nonumber\\
&\simeq {P(\tilde{t}_{\rm hub}(t, I)) \over \, a_1}{1\over I}.
\label{eq:derivePtI}
\end{align}
%%%%%%
Figures~3(c) and~\ref{fig:othernet}(d) and~\ref{fig:othernet}(h) suggest that for a given $t$, the number of infected nodes $I$ abruptly decreases from $I_2$ to $I_1$ with $I_2\gg I_1$ as $t_{\rm hub}$ increases in the region $t-\Delta \lesssim t_{\rm hub} \lesssim t$. The variation of $\tilde{t}_{\rm hub}(t, I)$ in the interval $I_1\lesssim I \lesssim I_2$ for a given $t$ is not larger than $\Delta$; $\tilde{t}_{\rm hub}(t, I_1)=t-(\ln I_1 -a_0)/a_1 \lesssim t$, and $\tilde{t}_{\rm hub}(t, I_2)=t-(\ln I_2 -a_0)/a_1 \gtrsim t-\Delta$. Since the width $w$ of $P(t_{\rm hub})$ is not smaller than $\Delta$, with $w\simeq10^4$ and $\Delta\simeq 5000$ as an example in the case of $\gamma=2.75$ and $N=10^6$, $P(\tilde{t}_{\rm hub}(t, I))$ is expected to vary only weakly with $I$ in the interval $I_1\lesssim I \lesssim I_2$ for a given $t$, allowing us to make the approximation $P(\tilde{t}_{\rm hub}(t, I))\simeq P_0$ with $P_0$ being a constant depending only on $t$ in Eq.~(\ref{eq:derivePtI}), reproducing Eq.~(3).
%%%%%%

\begin{figure}
\includegraphics[width=\columnwidth]{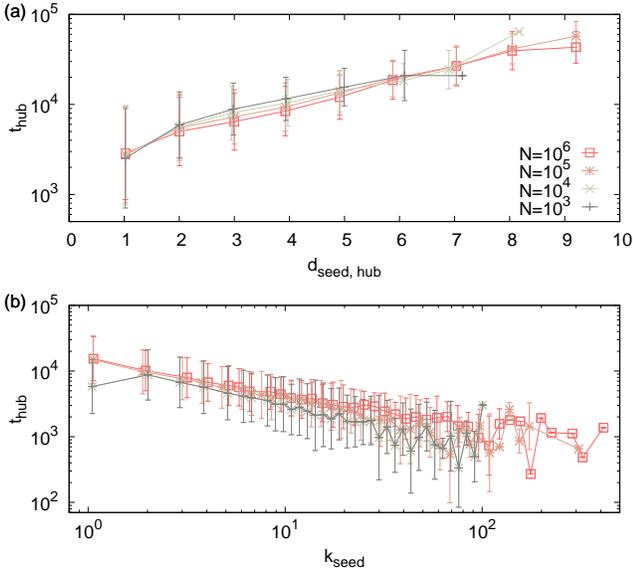}
\caption{The dependence of the hub-infection time on the initially infected node (seed). (a) The hub-infection time $t_{\rm hub}$  increases with the distance $d_{\rm seed, hub}$ of the seed node to the nearest hub node, having a degree larger than $0.3 k_{\rm max}$, in SF networks with $\gamma=2.75$ and different numbers of nodes $N$. (b) Plot of $t_{\rm hub}$ versus the degree  $k_{\rm seed}$ of the seed node. 
}
\label{fig:thubdeterminants}
\end{figure}

Actually, the asymptotic behavior $P_t(I)\sim I^{-1}$ is valid as long as the conditional probability $P_{t,t_{\rm hub}} (I)$ is well concentrated in the logarithmic scale at its mean $I_*= e^{\langle \ln I\rangle_{t,t_{\rm hub}}}$ with width $\Delta I$, for which we have approximately  $1=\int dI P_{t,t_{\rm hub}} (I) \simeq \Delta I P_{t,t_{\rm hub}} (I_*) \simeq {\Delta I\over \Delta t_{\rm hub}}\bigg|_{I=I_*} \int d\, t _{\rm hub}^{'} P_{t,t_{\rm hub}^{'}} (I_*)$, leading to  $\int d\, t _{\rm hub} P_{t,t_{\rm hub}} (I) \sim \bigg|{dt_{\rm hub} \over dI}\bigg|\sim I^{-1}$ from Eq.~(2). In Fig~\ref{fig:thubdeterminants}, the dependence of $t_{\rm hub}$ on the network characteristics of the seed is shown.

%%%%%%%% Figure: P(t03) %%%%%%%%%%%%%%%%%
%%%%%%%%%%%%%%%%%%%%%%%%%%%%%%%%%%

%%%%%%%% Figure: sigma KKhat %%%%%%%%%%%%%%%%%
\begin{figure}
\includegraphics[width=\columnwidth]{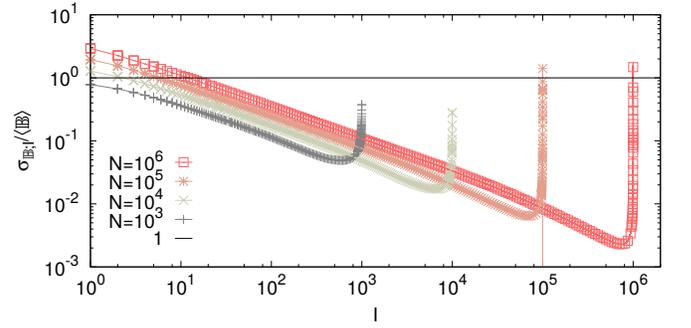}
\caption{The ratio of the standard deviation $\sigma_{\mathbb{B};I}$ to the mean $\langle \mathbb{B}\rangle_I$ of the number of boundary links for a given number of infected nodes $I$ in SF networks with $\gamma=2.75$. 
}
\label{fig:sigmaB}
\end{figure}
%%%%%%%%%%%%%%%%%%%%%%%%%%%%%%%%%%

\section{Derivation of Equation~(8)}
\label{app:pdki}
Suppose that there are $I$ infected nodes. Assuming that every link from the $N-I$ susceptible nodes is connected to one of the $I$ infected nodes with the same probability $q$, we find that a susceptible node reached from an infected node by a link has degree $k$ with probability
\begin{equation}
r_I(k) = {q\, k n(k|I) \over \sum_{k'} q k' n(k'|I)}  = {k  \over 2L - \Vmean_I} n(k|I),
\end{equation}
where $n(k|I)$ is the expected number of susceptible nodes of degree $k$ given $I$ infected nodes and $2L-\Vmean_I = \sum_k k n(k|I)$ is the expected number of links incident to susceptible nodes. It can be noticed that $q={\Bmean_I\over 2L-\Vmean_I}$, where $\Bmean_I$ is a well-defined value as seen in Fig.~\ref{fig:sigmaB}. When a new node is infected to become the $(I+1)$th infected node, the number of susceptible nodes of degree $k$ will be decreased by one if the newly infected node has degree $k$, which occurs with probability $r_I(k)$. Therefore, the decrease in the expected number of susceptible nodes of degree $k$, $n(k|I)-n(k|I+1)$, is equal to $r_I(k)$~\cite{PhysRevE.93.052310}:
$n(k|I)- n(k|I+1) = r_I(k) = { k \over 2L - \Vmean_I} n(k|I)$, or, equivalently,
\begin{equation}
n(k|I+1) = \left[1-{k \over 2L -\Vmean_I}\right] n(k|I).
\label{eq:nkrecursive}
\end{equation}
It holds that ${k\over 2L - \Vmean_I} \ll1$ for all $k$ in the intermediate stage, and therefore,  $n(k|I)$ is represented as
\begin{align}
n(k|I)=n(k|0)\prod_{I'=0}^{I-1}
\left( 1-\dfrac{k}{2L-\Vmean_I}\right)\simeq NP_{\rm degree}(k)e^{-{k\over \tilde{k}(I)}},
\label{eq:nki}
\end{align}
with the cut-off degree $\tilde{k}(I)$ depending on the number of infected nodes $I$ as 
\begin{align}
\tilde{k}(I)^{-1}&\equiv  \sum_{I'=0}^{I-1}{1\over 2L - \Vmean_{I'}} = {I\over 2L} \left(1+ 
%{1\over 2L } { \sum_{I'=0}^{I-1} K(I')\over I}
O\left({I\over 2L} \right)^{\min\{1,\gamma-2\}}\right)\nonumber\\
&\simeq {I\over 2L}
\label{eq:kci}
\end{align}
 and the initial condition  $n(k|0)=NP_{\rm degree}(k)$. The remainder in Eq.~(\ref{eq:kci}) can be evaluated by using Eq.~(\ref{eq:ki2}) derived below~\cite{PhysRevE.93.052310}. Using Eqs.~(\ref{eq:nki}) and (\ref{eq:kci}), we obtain Eq.~(8).

\section{Asymptotic behaviors of $\langle k_{\rm nn}\rangle(I)$}
\label{app:asymptotic}

In SF networks with an asymptotic power-law degree distribution $P_{\rm degree}(k) \simeq c_0 k^{-\gamma}$ for large $k$ with $c_0$ being a constant,  the numerator in Eq.~(8) may diverge with the smaller of the maximum degrees $k_{\rm max}$ and the cutoff $\tilde{k}(I)$ if the degree exponent $\gamma$ is smaller than $3$. If $\gamma>3$, the contribution of the summand for large $k$ is negligible and so is the effect of the exponential term in the whole sum, allowing us to approximate Eq.~(8) as 
\begin{equation}
\langle k_{\rm nn}\rangle(I) \simeq   \dfrac{ \sum_{k=1}^{k_{\rm max}}k^2 \, P_{\rm degree} (k) }{\langle k\rangle}  = \langle k_{\rm nn}\rangle
\label{eq:kinf_g3}
\end{equation}
with $\langle k\rangle = \sum_k k P_{\rm degree} (k)$. 
If $2<\gamma<3$, the slow decay of the summand for large $k$ causes the divergence of the numerator. From Eq.~(8), we find
\begin{align}
&\langle k_{\rm nn}\rangle(I) \simeq 
\dfrac{c_0 \sum_{k=1}^{k_{\rm max}}k^{2-\gamma}e^{-k/\tilde{k}(I)}}{\langle k\rangle} \nonumber \\
&\simeq  {c_0 \over \langle k\rangle}  \tilde{k}(I)^{3-\gamma}  \left[
 \Gamma\left(3-\gamma,{1\over \tilde{k}(I)}\right) - \Gamma\left(3-\gamma,{k_{\rm max}\over \tilde{k}(I)}\right)\right],
\label{eq:kinfexact}
\end{align}
where  we used the Euler-Maclaurin formula and the incomplete gamma function $\Gamma(s,z) \equiv \int_z^\infty dk \, k^{s-1} \, e^{-k}$. 
The values of $\langle k_{\rm nn}\rangle(I)$ evaluated with Eq.~(\ref{eq:kinfexact}) and by using the numerical degree distribution in Eq.~(8) are in good agreement. 
The incomplete gamma function is expanded as~\cite{incompletegammaftn}
\begin{align}
\Gamma(s,z) &\simeq  
\begin{cases}
\Gamma(s) - {z^s \over s} + O(z^{s+1}) & {\rm for} \, z\ll 1, \\
z^{s-1} e^{-z} &{\rm for} \, z\gg 1,
\end{cases}
\end{align} 
which leads us to find, for $2<\gamma<3$ in the limit $\tilde{k}(I), k_{\rm max}\to \infty$, the small- and large-$\tilde{k}(I)$ behaviors of $\langle k_{\rm nn}\rangle(I)$, given by 
\begin{align}
\langle k_{\rm nn}\rangle (I) &\simeq
\begin{cases}
{c_0\, k_{\rm max}^{3-\gamma} \over (3-\gamma) \langle k\rangle} & {\rm for}  \ \tilde{k}(I)\gg k_{\rm max},\\
{c_0 \,  \Gamma(3-\gamma) \tilde{k}(I)^{3-\gamma} \over \langle k\rangle} &{\rm for}  \ \tilde{k}(I)\ll k_{\rm max}.
\end{cases}
\label{eq:ki2}
\end{align}
Noting that $\langle k_{\rm nn}\rangle$ for $2<\gamma<3$ is evaluated as 
\begin{equation}
\langle k_{\rm{nn}} \rangle = {\sum_k k^2 P_d(k) \over\langle k\rangle} \simeq {c_0 \sum_{k=1}^{k_{\rm{max}}} k^{2-\gamma} \over\langle k\rangle} \simeq {c_0 \, k_{\rm max}^{3-\gamma} \over (3-\gamma) \langle k\rangle }
\end{equation}
and using $\tilde{k}(I) = 2L/I$ as in Eq.~(\ref{eq:kci}), 
we can rewrite Eq.~(\ref{eq:ki2}) as 
\begin{equation}
\langle k_{\rm nn} \rangle (I) \simeq
\begin{cases}
\langle k_{\rm nn}\rangle & {\rm for}   \ I\ll I_{\rm c},\\
\langle k_{\rm nn}\rangle \Gamma(4-\gamma) \left({I_{\rm c} \over I}\right)^{3-\gamma} &{\rm for}  \ I \gg I_{\rm c},
\end{cases}
\label{eq:finki}
\end{equation}
as in the main text. 

%\bibliography{/Users/dslee/Dropbox/2017SImodel/Manuscript/20190110_PREtransfer/20190226Reply/SI_pre_rev}

\begin{thebibliography}{36}%
\makeatletter
\providecommand \@ifxundefined [1]{%
 \@ifx{#1\undefined}
}%
\providecommand \@ifnum [1]{%
 \ifnum #1\expandafter \@firstoftwo
 \else \expandafter \@secondoftwo
 \fi
}%
\providecommand \@ifx [1]{%
 \ifx #1\expandafter \@firstoftwo
 \else \expandafter \@secondoftwo
 \fi
}%
\providecommand \natexlab [1]{#1}%
\providecommand \enquote  [1]{``#1''}%
\providecommand \bibnamefont  [1]{#1}%
\providecommand \bibfnamefont [1]{#1}%
\providecommand \citenamefont [1]{#1}%
\providecommand \href@noop [0]{\@secondoftwo}%
\providecommand \href [0]{\begingroup \@sanitize@url \@href}%
\providecommand \@href[1]{\@@startlink{#1}\@@href}%
\providecommand \@@href[1]{\endgroup#1\@@endlink}%
\providecommand \@sanitize@url [0]{\catcode `\\12\catcode `\$12\catcode
  `\&12\catcode `\#12\catcode `\^12\catcode `\_12\catcode `\%12\relax}%
\providecommand \@@startlink[1]{}%
\providecommand \@@endlink[0]{}%
\providecommand \url  [0]{\begingroup\@sanitize@url \@url }%
\providecommand \@url [1]{\endgroup\@href {#1}{\urlprefix }}%
\providecommand \urlprefix  [0]{URL }%
\providecommand \Eprint [0]{\href }%
\providecommand \doibase [0]{http://dx.doi.org/}%
\providecommand \selectlanguage [0]{\@gobble}%
\providecommand \bibinfo  [0]{\@secondoftwo}%
\providecommand \bibfield  [0]{\@secondoftwo}%
\providecommand \translation [1]{[#1]}%
\providecommand \BibitemOpen [0]{}%
\providecommand \bibitemStop [0]{}%
\providecommand \bibitemNoStop [0]{.\EOS\space}%
\providecommand \EOS [0]{\spacefactor3000\relax}%
\providecommand \BibitemShut  [1]{\csname bibitem#1\endcsname}%
\let\auto@bib@innerbib\@empty
%</preamble>
\bibitem [{\citenamefont {Barab\'{a}si}\ and\ \citenamefont
  {Albert}(1999)}]{barabasi99}%
  \BibitemOpen
  \bibfield  {author} {\bibinfo {author} {\bibfnamefont {A.-L.}\ \bibnamefont
  {Barab\'{a}si}}\ and\ \bibinfo {author} {\bibfnamefont {R.}~\bibnamefont
  {Albert}},\ }\href@noop {} {\bibfield  {journal} {\bibinfo  {journal}
  {Science}\ }\textbf {\bibinfo {volume} {286}},\ \bibinfo {pages} {509}
  (\bibinfo {year} {1999})}\BibitemShut {NoStop}%
\bibitem [{\citenamefont {Albert}\ \emph {et~al.}(2000)\citenamefont {Albert},
  \citenamefont {Jeong},\ and\ \citenamefont {Barab\'{a}si}}]{albert00}%
  \BibitemOpen
  \bibfield  {author} {\bibinfo {author} {\bibfnamefont {R.}~\bibnamefont
  {Albert}}, \bibinfo {author} {\bibfnamefont {H.}~\bibnamefont {Jeong}}, \
  and\ \bibinfo {author} {\bibfnamefont {A.-L.}\ \bibnamefont {Barab\'{a}si}},\
  }\href@noop {} {\bibfield  {journal} {\bibinfo  {journal} {Nature}\ }\textbf
  {\bibinfo {volume} {406}},\ \bibinfo {pages} {378} (\bibinfo {year}
  {2000})}\BibitemShut {NoStop}%
\bibitem [{\citenamefont {Pastor-Satorras}\ and\ \citenamefont
  {Vespignani}(2001{\natexlab{a}})}]{sis2}%
  \BibitemOpen
  \bibfield  {author} {\bibinfo {author} {\bibfnamefont {R.}~\bibnamefont
  {Pastor-Satorras}}\ and\ \bibinfo {author} {\bibfnamefont {A.}~\bibnamefont
  {Vespignani}},\ }\href@noop {} {\bibfield  {journal} {\bibinfo  {journal}
  {Phys. Rev. Lett.}\ }\textbf {\bibinfo {volume} {86}},\ \bibinfo {pages}
  {3200} (\bibinfo {year} {2001}{\natexlab{a}})}\BibitemShut {NoStop}%
\bibitem [{\citenamefont {Ichinomiya}(2004)}]{PhysRevE.70.026116}%
  \BibitemOpen
  \bibfield  {author} {\bibinfo {author} {\bibfnamefont {T.}~\bibnamefont
  {Ichinomiya}},\ }\href {\doibase 10.1103/PhysRevE.70.026116} {\bibfield
  {journal} {\bibinfo  {journal} {Phys. Rev. E}\ }\textbf {\bibinfo {volume}
  {70}},\ \bibinfo {pages} {026116} (\bibinfo {year} {2004})}\BibitemShut
  {NoStop}%
\bibitem [{\citenamefont {Dorogovtsev}\ \emph {et~al.}(2002)\citenamefont
  {Dorogovtsev}, \citenamefont {Goltsev},\ and\ \citenamefont
  {Mendes}}]{PhysRevE.66.016104}%
  \BibitemOpen
  \bibfield  {author} {\bibinfo {author} {\bibfnamefont {S.~N.}\ \bibnamefont
  {Dorogovtsev}}, \bibinfo {author} {\bibfnamefont {A.~V.}\ \bibnamefont
  {Goltsev}}, \ and\ \bibinfo {author} {\bibfnamefont {J.~F.~F.}\ \bibnamefont
  {Mendes}},\ }\href {\doibase 10.1103/PhysRevE.66.016104} {\bibfield
  {journal} {\bibinfo  {journal} {Phys. Rev. E}\ }\textbf {\bibinfo {volume}
  {66}},\ \bibinfo {pages} {016104} (\bibinfo {year} {2002})}\BibitemShut
  {NoStop}%
\bibitem [{\citenamefont {Igl\'oi}\ and\ \citenamefont
  {Turban}(2002)}]{PhysRevE.66.036140}%
  \BibitemOpen
  \bibfield  {author} {\bibinfo {author} {\bibfnamefont {F.}~\bibnamefont
  {Igl\'oi}}\ and\ \bibinfo {author} {\bibfnamefont {L.}~\bibnamefont
  {Turban}},\ }\href {\doibase 10.1103/PhysRevE.66.036140} {\bibfield
  {journal} {\bibinfo  {journal} {Phys. Rev. E}\ }\textbf {\bibinfo {volume}
  {66}},\ \bibinfo {pages} {036140} (\bibinfo {year} {2002})}\BibitemShut
  {NoStop}%
\bibitem [{\citenamefont {Lee}\ \emph {et~al.}(2004)\citenamefont {Lee},
  \citenamefont {Goh}, \citenamefont {Kahng},\ and\ \citenamefont
  {Kim}}]{lee04}%
  \BibitemOpen
  \bibfield  {author} {\bibinfo {author} {\bibfnamefont {D.-S.}\ \bibnamefont
  {Lee}}, \bibinfo {author} {\bibfnamefont {K.-I.}\ \bibnamefont {Goh}},
  \bibinfo {author} {\bibfnamefont {B.}~\bibnamefont {Kahng}}, \ and\ \bibinfo
  {author} {\bibfnamefont {D.}~\bibnamefont {Kim}},\ }\href@noop {} {\bibfield
  {journal} {\bibinfo  {journal} {Nucl. Phys. B}\ }\textbf {\bibinfo {volume}
  {696}},\ \bibinfo {pages} {351} (\bibinfo {year} {2004})}\BibitemShut
  {NoStop}%
\bibitem [{\citenamefont {Boccaletti}\ \emph {et~al.}(2006)\citenamefont
  {Boccaletti}, \citenamefont {Latora}, \citenamefont {Moreno}, \citenamefont
  {Chavez},\ and\ \citenamefont {Hwang}}]{Boccaletti:2006aa}%
  \BibitemOpen
  \bibfield  {author} {\bibinfo {author} {\bibfnamefont {S.}~\bibnamefont
  {Boccaletti}}, \bibinfo {author} {\bibfnamefont {V.}~\bibnamefont {Latora}},
  \bibinfo {author} {\bibfnamefont {Y.}~\bibnamefont {Moreno}}, \bibinfo
  {author} {\bibfnamefont {M.}~\bibnamefont {Chavez}}, \ and\ \bibinfo {author}
  {\bibfnamefont {D.~U.}\ \bibnamefont {Hwang}},\ }\href {\doibase
  https://doi.org/10.1016/j.physrep.2005.10.009} {\bibfield  {journal}
  {\bibinfo  {journal} {Phys. Rep.}\ }\textbf {\bibinfo {volume} {424}},\
  \bibinfo {pages} {175} (\bibinfo {year} {2006})}\BibitemShut {NoStop}%
\bibitem [{\citenamefont {Dorogovtsev}\ \emph {et~al.}(2008)\citenamefont
  {Dorogovtsev}, \citenamefont {Goltsev},\ and\ \citenamefont
  {Mendes}}]{dorogovtsev08}%
  \BibitemOpen
  \bibfield  {author} {\bibinfo {author} {\bibfnamefont {S.}~\bibnamefont
  {Dorogovtsev}}, \bibinfo {author} {\bibfnamefont {A.}~\bibnamefont
  {Goltsev}}, \ and\ \bibinfo {author} {\bibfnamefont {J.}~\bibnamefont
  {Mendes}},\ }\href@noop {} {\bibfield  {journal} {\bibinfo  {journal} {Rev.
  Mod. Phys.}\ }\textbf {\bibinfo {volume} {80}},\ \bibinfo {pages} {1275}
  (\bibinfo {year} {2008})}\BibitemShut {NoStop}%
\bibitem [{\citenamefont {Barrat}\ \emph {et~al.}(2008)\citenamefont {Barrat},
  \citenamefont {Barth{\'e}lemy},\ and\ \citenamefont
  {Vespignani}}]{barrat2008dynamical}%
  \BibitemOpen
  \bibfield  {author} {\bibinfo {author} {\bibfnamefont {A.}~\bibnamefont
  {Barrat}}, \bibinfo {author} {\bibfnamefont {M.}~\bibnamefont
  {Barth{\'e}lemy}}, \ and\ \bibinfo {author} {\bibfnamefont {A.}~\bibnamefont
  {Vespignani}},\ }\href {https://books.google.co.kr/books?id=TmgePn9uQD4C}
  {\emph {\bibinfo {title} {Dynamical Processes on Complex Networks}}}\
  (\bibinfo  {publisher} {Cambridge University Press},\ \bibinfo {address} {New
  York},\ \bibinfo {year} {2008})\BibitemShut {NoStop}%
\bibitem [{\citenamefont {Barth\'elemy}\ \emph {et~al.}(2004)\citenamefont
  {Barth\'elemy}, \citenamefont {Barrat}, \citenamefont {Pastor-Satorras},\
  and\ \citenamefont {Vespignani}}]{PhysRevLett.92.178701}%
  \BibitemOpen
  \bibfield  {author} {\bibinfo {author} {\bibfnamefont {M.}~\bibnamefont
  {Barth\'elemy}}, \bibinfo {author} {\bibfnamefont {A.}~\bibnamefont
  {Barrat}}, \bibinfo {author} {\bibfnamefont {R.}~\bibnamefont
  {Pastor-Satorras}}, \ and\ \bibinfo {author} {\bibfnamefont {A.}~\bibnamefont
  {Vespignani}},\ }\href {\doibase 10.1103/PhysRevLett.92.178701} {\bibfield
  {journal} {\bibinfo  {journal} {Phys. Rev. Lett.}\ }\textbf {\bibinfo
  {volume} {92}},\ \bibinfo {pages} {178701} (\bibinfo {year}
  {2004})}\BibitemShut {NoStop}%
\bibitem [{BAR(2005)}]{BARTHELEMY2005275}%
  \BibitemOpen
  \href {\doibase https://doi.org/10.1016/j.jtbi.2005.01.011} {\bibfield
  {journal} {\bibinfo  {journal} {J. Theor. Biol.}\ }\textbf {\bibinfo {volume}
  {235}},\ \bibinfo {pages} {275 } (\bibinfo {year} {2005})}\BibitemShut
  {NoStop}%
\bibitem [{\citenamefont {Volz}(2008)}]{Volz2008}%
  \BibitemOpen
  \bibfield  {author} {\bibinfo {author} {\bibfnamefont {E.}~\bibnamefont
  {Volz}},\ }\href {\doibase 10.1007/s00285-007-0116-4} {\bibfield  {journal}
  {\bibinfo  {journal} {J. Math. Biol.}\ }\textbf {\bibinfo {volume} {56}},\
  \bibinfo {pages} {293} (\bibinfo {year} {2008})}\BibitemShut {NoStop}%
\bibitem [{\citenamefont {Miller}(2011)}]{Miller2011}%
  \BibitemOpen
  \bibfield  {author} {\bibinfo {author} {\bibfnamefont {J.~C.}\ \bibnamefont
  {Miller}},\ }\href {\doibase 10.1007/s00285-010-0337-9} {\bibfield  {journal}
  {\bibinfo  {journal} {J. Math. Biol.}\ }\textbf {\bibinfo {volume} {62}},\
  \bibinfo {pages} {349} (\bibinfo {year} {2011})}\BibitemShut {NoStop}%
\bibitem [{\citenamefont {Gleeson}(2013)}]{PhysRevX.3.021004}%
  \BibitemOpen
  \bibfield  {author} {\bibinfo {author} {\bibfnamefont {J.~P.}\ \bibnamefont
  {Gleeson}},\ }\href {\doibase 10.1103/PhysRevX.3.021004} {\bibfield
  {journal} {\bibinfo  {journal} {Phys. Rev. X}\ }\textbf {\bibinfo {volume}
  {3}},\ \bibinfo {pages} {021004} (\bibinfo {year} {2013})}\BibitemShut
  {NoStop}%
\bibitem [{\citenamefont {Mata}\ and\ \citenamefont
  {Ferreira}(2013)}]{Mata_2013}%
  \BibitemOpen
  \bibfield  {author} {\bibinfo {author} {\bibfnamefont {A.~S.}\ \bibnamefont
  {Mata}}\ and\ \bibinfo {author} {\bibfnamefont {S.~C.}\ \bibnamefont
  {Ferreira}},\ }\href {\doibase 10.1209/0295-5075/103/48003} {\bibfield
  {journal} {\bibinfo  {journal} {EPL}\ }\textbf {\bibinfo {volume} {103}},\
  \bibinfo {pages} {48003} (\bibinfo {year} {2013})}\BibitemShut {NoStop}%
\bibitem [{\citenamefont {Trapman}(2007)}]{TRAPMAN2007464}%
  \BibitemOpen
  \bibfield  {author} {\bibinfo {author} {\bibfnamefont {P.}~\bibnamefont
  {Trapman}},\ }\href {\doibase https://doi.org/10.1016/j.mbs.2007.05.011}
  {\bibfield  {journal} {\bibinfo  {journal} {Math. Biosci.}\ }\textbf
  {\bibinfo {volume} {210}},\ \bibinfo {pages} {464 } (\bibinfo {year}
  {2007})}\BibitemShut {NoStop}%
\bibitem [{\citenamefont {Vazquez}(2006)}]{PhysRevLett.96.038702}%
  \BibitemOpen
  \bibfield  {author} {\bibinfo {author} {\bibfnamefont {A.}~\bibnamefont
  {Vazquez}},\ }\href {\doibase 10.1103/PhysRevLett.96.038702} {\bibfield
  {journal} {\bibinfo  {journal} {Phys. Rev. Lett.}\ }\textbf {\bibinfo
  {volume} {96}},\ \bibinfo {pages} {038702} (\bibinfo {year}
  {2006})}\BibitemShut {NoStop}%
\bibitem [{\citenamefont {Krishnarajah}\ \emph {et~al.}(2005)\citenamefont
  {Krishnarajah}, \citenamefont {Cook}, \citenamefont {Marion},\ and\
  \citenamefont {Gibson}}]{Krishnarajah2005}%
  \BibitemOpen
  \bibfield  {author} {\bibinfo {author} {\bibfnamefont {I.}~\bibnamefont
  {Krishnarajah}}, \bibinfo {author} {\bibfnamefont {A.}~\bibnamefont {Cook}},
  \bibinfo {author} {\bibfnamefont {G.}~\bibnamefont {Marion}}, \ and\ \bibinfo
  {author} {\bibfnamefont {G.}~\bibnamefont {Gibson}},\ }\href {\doibase
  10.1016/j.bulm.2004.11.002} {\bibfield  {journal} {\bibinfo  {journal} {B.
  Math. Biol.}\ }\textbf {\bibinfo {volume} {67}},\ \bibinfo {pages} {855}
  (\bibinfo {year} {2005})}\BibitemShut {NoStop}%
\bibitem [{\citenamefont {Bauch}(2005)}]{BAUCH2005217}%
  \BibitemOpen
  \bibfield  {author} {\bibinfo {author} {\bibfnamefont {C.~T.}\ \bibnamefont
  {Bauch}},\ }\href {\doibase https://doi.org/10.1016/j.mbs.2005.06.005}
  {\bibfield  {journal} {\bibinfo  {journal} {Math. Biosci.}\ }\textbf
  {\bibinfo {volume} {198}},\ \bibinfo {pages} {217 } (\bibinfo {year}
  {2005})}\BibitemShut {NoStop}%
\bibitem [{\citenamefont {Peyrard}\ \emph {et~al.}(2008)\citenamefont
  {Peyrard}, \citenamefont {Dieckmann},\ and\ \citenamefont
  {Franc}}]{PEYRARD2008383}%
  \BibitemOpen
  \bibfield  {author} {\bibinfo {author} {\bibfnamefont {N.}~\bibnamefont
  {Peyrard}}, \bibinfo {author} {\bibfnamefont {U.}~\bibnamefont {Dieckmann}},
  \ and\ \bibinfo {author} {\bibfnamefont {A.}~\bibnamefont {Franc}},\ }\href
  {\doibase https://doi.org/10.1016/j.tpb.2007.12.006} {\bibfield  {journal}
  {\bibinfo  {journal} {Theor. Popul. Biol.}\ }\textbf {\bibinfo {volume}
  {73}},\ \bibinfo {pages} {383 } (\bibinfo {year} {2008})}\BibitemShut
  {NoStop}%
\bibitem [{\citenamefont {Karrer}\ and\ \citenamefont
  {Newman}(2010)}]{PhysRevE.82.016101}%
  \BibitemOpen
  \bibfield  {author} {\bibinfo {author} {\bibfnamefont {B.}~\bibnamefont
  {Karrer}}\ and\ \bibinfo {author} {\bibfnamefont {M.~E.~J.}\ \bibnamefont
  {Newman}},\ }\href {\doibase 10.1103/PhysRevE.82.016101} {\bibfield
  {journal} {\bibinfo  {journal} {Phys. Rev. E}\ }\textbf {\bibinfo {volume}
  {82}},\ \bibinfo {pages} {016101} (\bibinfo {year} {2010})}\BibitemShut
  {NoStop}%
\bibitem [{\citenamefont {Altarelli}\ \emph {et~al.}(2014)\citenamefont
  {Altarelli}, \citenamefont {Braunstein}, \citenamefont {Dall'Asta},
  \citenamefont {Lage-Castellanos},\ and\ \citenamefont
  {Zecchina}}]{PhysRevLett.112.118701}%
  \BibitemOpen
  \bibfield  {author} {\bibinfo {author} {\bibfnamefont {F.}~\bibnamefont
  {Altarelli}}, \bibinfo {author} {\bibfnamefont {A.}~\bibnamefont
  {Braunstein}}, \bibinfo {author} {\bibfnamefont {L.}~\bibnamefont
  {Dall'Asta}}, \bibinfo {author} {\bibfnamefont {A.}~\bibnamefont
  {Lage-Castellanos}}, \ and\ \bibinfo {author} {\bibfnamefont
  {R.}~\bibnamefont {Zecchina}},\ }\href {\doibase
  10.1103/PhysRevLett.112.118701} {\bibfield  {journal} {\bibinfo  {journal}
  {Phys. Rev. Lett.}\ }\textbf {\bibinfo {volume} {112}},\ \bibinfo {pages}
  {118701} (\bibinfo {year} {2014})}\BibitemShut {NoStop}%
\bibitem [{\citenamefont {Catanzaro}\ \emph {et~al.}(2005)\citenamefont
  {Catanzaro}, \citenamefont {Bogu{\~n}{\'a}},\ and\ \citenamefont
  {Pastor-Satorras}}]{ucm}%
  \BibitemOpen
  \bibfield  {author} {\bibinfo {author} {\bibfnamefont {M.}~\bibnamefont
  {Catanzaro}}, \bibinfo {author} {\bibfnamefont {M.}~\bibnamefont
  {Bogu{\~n}{\'a}}}, \ and\ \bibinfo {author} {\bibfnamefont {R.}~\bibnamefont
  {Pastor-Satorras}},\ }\href@noop {} {\bibfield  {journal} {\bibinfo
  {journal} {Phys. Rev. E}\ }\textbf {\bibinfo {volume} {71}},\ \bibinfo
  {pages} {027103} (\bibinfo {year} {2005})}\BibitemShut {NoStop}%
\bibitem [{\citenamefont {Kim}\ \emph {et~al.}(2019)\citenamefont {Kim},
  \citenamefont {Lee}, \citenamefont {Barbier}, \citenamefont {Choi},
  \citenamefont {Kim}, \citenamefont {Yoo},\ and\ \citenamefont {Lee}}]{hk}%
  \BibitemOpen
  \bibfield  {author} {\bibinfo {author} {\bibfnamefont {H.~K.}\ \bibnamefont
  {Kim}}, \bibinfo {author} {\bibfnamefont {M.~J.}\ \bibnamefont {Lee}},
  \bibinfo {author} {\bibfnamefont {M.}~\bibnamefont {Barbier}}, \bibinfo
  {author} {\bibfnamefont {S.-G.}\ \bibnamefont {Choi}}, \bibinfo {author}
  {\bibfnamefont {M.~S.}\ \bibnamefont {Kim}}, \bibinfo {author} {\bibfnamefont
  {H.-H.}\ \bibnamefont {Yoo}}, \ and\ \bibinfo {author} {\bibfnamefont
  {D.-S.}\ \bibnamefont {Lee}},\ }\href@noop {} {\bibfield  {journal} {\bibinfo
   {journal} {(unpublished)}} }\BibitemShut {NoStop}%
\bibitem [{\citenamefont {Porter}\ and\ \citenamefont
  {Gleeson}(2016)}]{portergleeson2016}%
  \BibitemOpen
  \bibfield  {author} {\bibinfo {author} {\bibfnamefont {M.~A.}\ \bibnamefont
  {Porter}}\ and\ \bibinfo {author} {\bibfnamefont {J.~P.}\ \bibnamefont
  {Gleeson}},\ }\href@noop {} {\emph {\bibinfo {title} {Dynamical Systems on
  Networks}}}\ (\bibinfo  {publisher} {Springer},\ \bibinfo {year}
  {2016})\BibitemShut {NoStop}%
\bibitem [{\citenamefont {Kitsak}\ \emph {et~al.}(2010)\citenamefont {Kitsak},
  \citenamefont {Gallos}, \citenamefont {Havlin}, \citenamefont {Liljeros},
  \citenamefont {Muchnik}, \citenamefont {Stanley},\ and\ \citenamefont
  {Makse}}]{Kitsak:2010aa}%
  \BibitemOpen
  \bibfield  {author} {\bibinfo {author} {\bibfnamefont {M.}~\bibnamefont
  {Kitsak}}, \bibinfo {author} {\bibfnamefont {L.~K.}\ \bibnamefont {Gallos}},
  \bibinfo {author} {\bibfnamefont {S.}~\bibnamefont {Havlin}}, \bibinfo
  {author} {\bibfnamefont {F.}~\bibnamefont {Liljeros}}, \bibinfo {author}
  {\bibfnamefont {L.}~\bibnamefont {Muchnik}}, \bibinfo {author} {\bibfnamefont
  {H.~E.}\ \bibnamefont {Stanley}}, \ and\ \bibinfo {author} {\bibfnamefont
  {H.~A.}\ \bibnamefont {Makse}},\ }\href {https://doi.org/10.1038/nphys1746}
  {\bibfield  {journal} {\bibinfo  {journal} {Nat. Phys.}\ }\textbf {\bibinfo
  {volume} {6}},\ \bibinfo {pages} {888} (\bibinfo {year} {2010})}\BibitemShut
  {NoStop}%
\bibitem [{\citenamefont {Massaro}\ \emph {et~al.}(2018)\citenamefont
  {Massaro}, \citenamefont {Ganin}, \citenamefont {Perra}, \citenamefont
  {Linkov},\ and\ \citenamefont {Vespignani}}]{Massaro:2018aa}%
  \BibitemOpen
  \bibfield  {author} {\bibinfo {author} {\bibfnamefont {E.}~\bibnamefont
  {Massaro}}, \bibinfo {author} {\bibfnamefont {A.}~\bibnamefont {Ganin}},
  \bibinfo {author} {\bibfnamefont {N.}~\bibnamefont {Perra}}, \bibinfo
  {author} {\bibfnamefont {I.}~\bibnamefont {Linkov}}, \ and\ \bibinfo {author}
  {\bibfnamefont {A.}~\bibnamefont {Vespignani}},\ }\href {\doibase
  10.1038/s41598-018-19706-2} {\bibfield  {journal} {\bibinfo  {journal} {Sci.
  Rep.}\ }\textbf {\bibinfo {volume} {8}},\ \bibinfo {pages} {1859} (\bibinfo
  {year} {2018})}\BibitemShut {NoStop}%
\bibitem [{\citenamefont {Morone}\ and\ \citenamefont
  {Makse}(2015)}]{Morone:2015aa}%
  \BibitemOpen
  \bibfield  {author} {\bibinfo {author} {\bibfnamefont {F.}~\bibnamefont
  {Morone}}\ and\ \bibinfo {author} {\bibfnamefont {H.~A.}\ \bibnamefont
  {Makse}},\ }\href {https://doi.org/10.1038/nature14604} {\bibfield  {journal}
  {\bibinfo  {journal} {Nature}\ }\textbf {\bibinfo {volume} {524}},\ \bibinfo
  {pages} {65} (\bibinfo {year} {2015})}\BibitemShut {NoStop}%
\bibitem [{\citenamefont {Radicchi}\ and\ \citenamefont
  {Castellano}(2017)}]{PhysRevE.95.012318}%
  \BibitemOpen
  \bibfield  {author} {\bibinfo {author} {\bibfnamefont {F.}~\bibnamefont
  {Radicchi}}\ and\ \bibinfo {author} {\bibfnamefont {C.}~\bibnamefont
  {Castellano}},\ }\href {\doibase 10.1103/PhysRevE.95.012318} {\bibfield
  {journal} {\bibinfo  {journal} {Phys. Rev. E}\ }\textbf {\bibinfo {volume}
  {95}},\ \bibinfo {pages} {012318} (\bibinfo {year} {2017})}\BibitemShut
  {NoStop}%
\bibitem [{\citenamefont {Pastor-Satorras}\ and\ \citenamefont
  {Vespignani}(2001{\natexlab{b}})}]{sis1}%
  \BibitemOpen
  \bibfield  {author} {\bibinfo {author} {\bibfnamefont {R.}~\bibnamefont
  {Pastor-Satorras}}\ and\ \bibinfo {author} {\bibfnamefont {A.}~\bibnamefont
  {Vespignani}},\ }\href@noop {} {\bibfield  {journal} {\bibinfo  {journal}
  {Phys. Rev. E}\ }\textbf {\bibinfo {volume} {63}},\ \bibinfo {pages} {066117}
  (\bibinfo {year} {2001}{\natexlab{b}})}\BibitemShut {NoStop}%
\bibitem [{\citenamefont {Pastor-Satorras}\ \emph {et~al.}(2015)\citenamefont
  {Pastor-Satorras}, \citenamefont {Castellano}, \citenamefont {Van~Mieghem},\
  and\ \citenamefont {Vespignani}}]{RevModPhys.87.925}%
  \BibitemOpen
  \bibfield  {author} {\bibinfo {author} {\bibfnamefont {R.}~\bibnamefont
  {Pastor-Satorras}}, \bibinfo {author} {\bibfnamefont {C.}~\bibnamefont
  {Castellano}}, \bibinfo {author} {\bibfnamefont {P.}~\bibnamefont
  {Van~Mieghem}}, \ and\ \bibinfo {author} {\bibfnamefont {A.}~\bibnamefont
  {Vespignani}},\ }\href {\doibase 10.1103/RevModPhys.87.925} {\bibfield
  {journal} {\bibinfo  {journal} {Rev. Mod. Phys.}\ }\textbf {\bibinfo {volume}
  {87}},\ \bibinfo {pages} {925} (\bibinfo {year} {2015})}\BibitemShut
  {NoStop}%
\bibitem [{\citenamefont {Devriendt}\ and\ \citenamefont
  {Van~Mieghem}(2017)}]{PhysRevE.96.052314}%
  \BibitemOpen
  \bibfield  {author} {\bibinfo {author} {\bibfnamefont {K.}~\bibnamefont
  {Devriendt}}\ and\ \bibinfo {author} {\bibfnamefont {P.}~\bibnamefont
  {Van~Mieghem}},\ }\href {\doibase 10.1103/PhysRevE.96.052314} {\bibfield
  {journal} {\bibinfo  {journal} {Phys. Rev. E}\ }\textbf {\bibinfo {volume}
  {96}},\ \bibinfo {pages} {052314} (\bibinfo {year} {2017})}\BibitemShut
  {NoStop}%
\bibitem [{\citenamefont {Miller}\ \emph {et~al.}(2012)\citenamefont {Miller},
  \citenamefont {Slim},\ and\ \citenamefont {Volz}}]{C.:2012aa}%
  \BibitemOpen
  \bibfield  {author} {\bibinfo {author} {\bibfnamefont {J.~C.}\ \bibnamefont
  {Miller}}, \bibinfo {author} {\bibfnamefont {A.~C.}\ \bibnamefont {Slim}}, \
  and\ \bibinfo {author} {\bibfnamefont {E.~M.}\ \bibnamefont {Volz}},\ }\href
  {\doibase 10.1098/rsif.2011.0403} {\bibfield  {journal} {\bibinfo  {journal}
  {J. Roy. Soc. Interface}\ }\textbf {\bibinfo {volume} {9}},\ \bibinfo {pages}
  {890} (\bibinfo {year} {2012})}\BibitemShut {NoStop}%
\bibitem [{\citenamefont {Kim}\ \emph {et~al.}(2016)\citenamefont {Kim},
  \citenamefont {Kyoung},\ and\ \citenamefont {Lee}}]{PhysRevE.93.052310}%
  \BibitemOpen
  \bibfield  {author} {\bibinfo {author} {\bibfnamefont {K.}~\bibnamefont
  {Kim}}, \bibinfo {author} {\bibfnamefont {J.}~\bibnamefont {Kyoung}}, \ and\
  \bibinfo {author} {\bibfnamefont {D.-S.}\ \bibnamefont {Lee}},\ }\href
  {\doibase 10.1103/PhysRevE.93.052310} {\bibfield  {journal} {\bibinfo
  {journal} {Phys. Rev. E}\ }\textbf {\bibinfo {volume} {93}},\ \bibinfo
  {pages} {052310} (\bibinfo {year} {2016})}\BibitemShut {NoStop}%
\bibitem [{inc()}]{incompletegammaftn}%
  \BibitemOpen
  \href {http://functions.wolfram.com/GammaBetaErf/Gamma2/} {{\bibinfo
  {title} {Mathematical Functions Site}}},\ \bibinfo {address}
  {http://functions.wolfram.com/GammaBetaErf/Gamma2}\BibitemShut {NoStop}%
\end{thebibliography}


\end{document}